\documentclass[reprint,
nofootinbib,
amssymb,
aps,
unsortedaddress,
]{revtex4-1}

\usepackage[intlimits]{amsmath}
\usepackage{accents}

\usepackage{graphicx}
\usepackage{dcolumn}
\usepackage{bm}
\usepackage[english]{babel}

\usepackage{tikz}
\usetikzlibrary{arrows.meta}
\usepackage[compat=1.1.0]{tikz-feynman}

\usepackage{verbatim}

\usepackage[colorlinks=true, linkcolor = black,
urlcolor  = black,
citecolor = black]{hyperref}

\begin{document}

\title{Inflaton-driven early dark energy}

\author{\href{https://sites.google.com/site/maziashvili/}{Michael~Maziashvili}}
\email{maziashvili@iliauni.edu.ge}

\affiliation{\vspace{0.2cm} \href{https://sites.google.com/site/maziashvili/}{School of Natural Sciences and Medicine, Ilia State University, \\ 3/5 Cholokashvili Ave., Tbilisi 0162, Georgia}}


\begin{abstract}  
	
By arranging the control parameters, we examine whether the mass varying neutrino model PRD 103, 063540 (2021), enabling one to unify inflation with the present dark energy, can be used for producing an early dark energy. The model works in the following way. At early stages of the Big-Bang, the inflaton trapped in the minimum at $\phi = 0$ gets uplifted due to interaction with neutrinos and starts to roll down to one of the degenerate minima of the effective potential and after a while gets anchored at this minimum, which in turn evolves in time very slowly. Correspondingly, the early dark energy taking place as a result of this dynamical symmetry breaking also varies in time very slowly. Shortly before the recombination epoch, however, the symmetry is restored and early dark energy disappears. A typical problem of the mass varying neutrino models is that they can hardly provide the needed amount of early dark energy at the tree-level because of smallness of neutrino masses. Nevertheless, the quantum fluctuations of $\phi$ can do the job in providing sufficient early dark energy under assumption that inflationary energy scale is of the order of $1$\,TeV. Radiative as well as thermal corrections coming from the neutrino sector do not affect the model significantly. As for the gravity induced corrections to the effective potential - they can be safely ignored.



\end{abstract}

\pacs{Valid PACS appear here}
\maketitle

\section{Briefly about the early dark energy}

The direct measurement of the "present" Hubble parameter by means of the supernova redshift observations tels us that \cite{Riess:2019cxk, Pesce:2020xfe, Freedman:2020dne} $H_0 \simeq 74$ km s$^{-1}$Mpc$^{-1}$. On the other hand, the locations of the acoustic peaks in CMB spectrum (measured with a very high accuracy) that determine the angular size of the sound horizon at recombination, 

\begin{eqnarray}
\theta_* = \frac{r_s(z_*)}{d_A(z_*)} ~, \nonumber 
\end{eqnarray} in conjunction with the $\Lambda$CDM-based evaluation of $r_s(z_*)$, allows one to evaluate the present value of Hubble parameter as $H_0 \simeq 67$ km s$^{-1}$Mpc$^{-1}$ \cite{Addison:2017fdm, Planck:2018vyg}. One of the possibilities to resolve or somewhat alleviate this mismatch is the early dark energy increasing the expansion rate in the early universe prior to recombination and thus reducing the sound horizon

\begin{eqnarray}
r_s(z_*) = \int_{z_*}^\infty\mathrm{d}z \, \frac{c_s(z)}{H(z)} ~, \nonumber 
\end{eqnarray} where $c_s(z)$ is the speed of sound in the baryon-photon fluid. It automatically requires the reduction of comoving angular diameter distance

\begin{eqnarray}
d_A(z_*) = \int_0^{z_*}\mathrm{d}z \, \frac{1}{H(z)} ~, \nonumber 
\end{eqnarray} since $\theta_*$ is fixed and thus results in a bigger value of the present Hubble parameter: $H_0 \sim \theta_*/r_s$.

The well known EDE models \cite{Poulin:2018dzj, Agrawal:2019lmo, Smith:2019ihp}, are built by introducing a scalar field which at very early times is supposed to be displaced from its minimum. The field motion is supposed to be negligible at first because of the Hubble drag that results in the cosmological constant providing an insignificant contribution to the total energy density of the universe in the past ($\rho_\phi \ll \rho_{tot}$). But around the matter-radiation equality, the ratio $\rho_\phi/\rho_{tot}$ approaches the value $0.1$ allowing one to resolve the Hubble tension problem. Around this time it is supposed that the Hubble parameter falls below the effective mass of the field and the field starts to move towards a minimum of the potential. Besides, the model to work, it is required that EDE redshifts away at least as fast as radiation shortly after the matter-radiation equality \cite{Poulin:2018cxd}. The parameters of such models are initial conditions, which are essential for providing a sufficient amount of $\rho_\phi$, and an extremely small mass scale, required to achieve kination at the time of recombination. Thus, in general, such models need to be fine tuned. To mitigate some of the fine tuning problems, it was suggested in \cite{Sakstein:2019fmf, CarrilloGonzalez:2020oac} to consider the coupling of EDE scalar field with the cosmic neutrino background (CNB) for dynamical introduction of a time-scale at which the fractional EDE peaks - by means of the temperature scale at which the CNB starts to transit to the non-relativistic regime: $T_\nu \simeq m_\nu$. This sort of models have been studied extensively in the context of dark energy (DE) \cite{Fardon:2003eh, Peccei:2004sz, Wetterich:2007kr, Amendola:2007yx, Brookfield:2005bz, Brookfield:2005bz} to address the coincidence problem. Motivated by the papers \cite{Sakstein:2019fmf, CarrilloGonzalez:2020oac}, we want to explore here the properties of a particular model of this kind enabling one one to use the same scalar field for DE and inflation \cite{Kepuladze:2021tsb} - to see if one can develop a similar model unifying EDE scalar field with the inflaton. This model admits $\mathcal{Z}_2$ symmetry, which breaks down in the course of cosmological evolution leading to the appearance of DE \cite{Pietroni:2005pv, MohseniSadjadi:2017pne}. At later times, this symmetry gets restored marking the disappearance of DE. Here we are talking about the second order phase transition. In particular, the field trapped initially in the minimum at $\phi = 0$, gets uplifted due to coupling with neutrinos  and rolls down to one of the minima of the effective potential leading thereby to the dynamical breaking of symmetry and correspondingly to the emergence of EDE. In contrast to the DE, the particular model we want to discuss does not provide the needed amount of EDE at the tree-level because of smallness of neutrino masses. It is worth noting that similar observation is made in \cite{Sakstein:2019fmf} as well\footnote{See the paragraph on the 4th page of arXive version of this paper, which starts with the sentence "Interestingly, our proposal is on the verge of being excluded, and may even be so already".}. However, in the present model EDE gets significantly amplified by the quantum fluctuations of $\phi$, which may be sufficient for resolving the Hubble tension problem. The EDE produced this way decays with CNB temperature as $\ln (T_\nu/T_\nu^*)$ and remains subdominant until the epoch of matter-radiation equality. Around the time of matter-radiation equality its existence becomes significant but shortly after this, at the temperature $T_\nu^*$, the field reaches minimum at $\phi=0$ (marking the restoration of $\mathcal{Z}_2$ symmetry) and EDE disappears.

\section{The model at the tree-level}

We start by discussing the model which has already been explored in some detail in \cite{Kepuladze:2021tsb}. This model, obtained by a slight reformulation of the model discussed previously in \cite{MohseniSadjadi:2017pne}, consists of the $\mathcal{Z}_2$ symmetric potential and $\phi$-$\nu$ coupling of the form

\begin{eqnarray}\label{example1}
U(\phi) = V\left(1-\mathrm{e}^{-\alpha \phi^2/M^2_P}\right) ~, ~ m_\nu(\phi) = \mu_\nu\mathrm{e}^{-\beta \phi^2/M^2_P} ~. ~~~
\end{eqnarray} The parameters

\begin{eqnarray}
\alpha \simeq 6.4 ~, ~~ 200 \,\text{MeV} \lesssim V^{1/4}  \lesssim 10^{13} \, \text{TeV} ~, \nonumber 
\end{eqnarray} are determined by the requirement to have a successful inflationary scenario \cite{Kepuladze:2021tsb}. For definiteness, let us consider a quasi-degenerate spectrum of neutrino masses (see sections 14 and 26 in \cite{ParticleDataGroup:2020ssz})

\begin{eqnarray}
m^0_1 \simeq m^0_2 \simeq m^0_3 \simeq 0.05 \, \text{eV} ~. \nonumber 
\end{eqnarray} The parameter $\mu_\nu$ in Eq.\eqref{example1} is set by this mass scale

\begin{eqnarray}\label{massscale}
	\mu_\nu \simeq 0.05 \, \text{eV} ~. 
\end{eqnarray} There still remains two free parameters $V$ and $\beta$ that we can use for constructing a working unified model of EDE and inflaton. As we shall see shortly, in the post-inflationary epochs the value of $\phi$ is very small as compared to $M_P$. More precisely

\begin{eqnarray}\label{smalness}
	(\phi /M_P)^2 \ll 1 ~, 
\end{eqnarray} and one can approximate the potential by

\begin{eqnarray}\label{example2potential}
	U(\phi) \approx \frac{\alpha V \phi^2}{M_P^2} ~. 
\end{eqnarray} The equations of motion for the system read \cite{Kepuladze:2021tsb}

\begin{eqnarray}
&& \dot{\rho}_\nu + 3H(\rho_\nu +p_\nu) =  \frac{\mathrm{d}\ln m_\nu}{\mathrm{d}\phi}(\rho_\nu - 3p_\nu)\dot{\phi} ~, \label{continuity} \\ && 
\ddot{\phi} +3H\dot{\phi} +U'(\phi) = -\frac{\mathrm{d}\ln m_\nu}{\mathrm{d}\phi} (\rho_\nu - 3p_\nu) ~, \label{eqofmot} \\ && H^2 = \frac{8\pi}{3M_P^2}(\rho_\phi +\rho_r+\rho_m) \label{Friedmann}~.
\end{eqnarray} First let us evaluate the quantity $ \rho_\nu - 3p_\nu $ entering the Eqs.(\ref{continuity}, \ref{eqofmot}). For free streaming neutrinos at a given temperature $T_\nu (\propto a^{-1})$, one finds that (see for instance \cite{Kepuladze:2021tsb}) 

\begin{eqnarray}
\frac{\mathrm{d}\ln m_\nu}{\mathrm{d}\phi} (\rho_\nu - 3p_\nu) = \frac{\partial\rho_\nu(\phi, T_\nu)}{\partial \phi} ~, \nonumber 
\end{eqnarray} and correspondingly the Eq.\eqref{eqofmot} takes the form

\begin{eqnarray}\label{scalar}
\ddot{\phi} +3H\dot{\phi} = -\partial_\phi \mathfrak{U}(\phi, T_\nu) ~, 
\end{eqnarray} where the tree-level effective potential

\begin{eqnarray}\label{tree-level}
\mathfrak{U}(\phi, T_\nu) \equiv  U(\phi) + \rho_\nu(\phi, T_\nu) ~. 
\end{eqnarray} Let us note that this potential is not defined uniquely, it can be replaced for instance by    

\begin{eqnarray}\label{effectivepotential2}
\widetilde{\mathfrak{U}}(\phi, T_\nu) = U(\phi) + \rho_\nu(\phi, T_\nu) - \rho_\nu(0, T_\nu) ~. \nonumber 
\end{eqnarray} We assume now that the temperature of CNB is smaller than its decoupling temperature ($\simeq 2$ MeV) so that one can use the phase space distribution function of free streaming neutrinos

\begin{eqnarray} \label{simkvrive} 
&&\rho_\nu = \frac{3\mathsf{g}}{a^3}\int\frac{\mathrm{d}^3k}{(2\pi)^3}\, \frac{\varepsilon_\nu(\mathbf{k})}{\mathrm{e}^{k/aT_\nu} + 1} ~,  ~   \varepsilon_\nu(\mathbf{k}) = \sqrt{\frac{\mathbf{k}^2}{a^2}+m_\nu^2} ~,  \nonumber \\&& p_\nu  =  \frac{\mathsf{3g}}{3a^5}\int\frac{\mathrm{d}^3k}{(2\pi)^3}\frac{k^2}{\varepsilon_\nu(\mathbf{k})\Big(\mathrm{e}^{k/aT_\nu} + 1\Big)}  ~, 
\end{eqnarray} where $\mathsf{g}( = 2)$ denotes the two helicity states per flavor, that is, the expression (\ref{simkvrive}) counts both neutrinos and anti-neutrinos, and the additional factor $3$ stands for the (effective) number of neutrino species \cite{deSalas:2016ztq}. The temperature of the CNB can be parameterized as

\begin{eqnarray}\label{temperature}
	T_\nu = 1.7\times 10^{-4} (1+z) \, \text{eV}  ~. 
\end{eqnarray} Let us first work out the asymptotic regime $\mu_\nu/T_\nu \ll 1$, 

\begin{eqnarray}\label{largeT}
\rho_\nu = \frac{3T_\nu^4}{\pi^2} \int_0^\infty\mathrm{d}\xi \, \frac{\xi^2\sqrt{\xi^2 +m_\nu^2/T_\nu^2}}{\mathrm{e}^\xi +1} \to   \frac{7\pi^2 T_\nu^4}{40} + \frac{m_\nu^2 T_\nu^2}{8} ~. ~~~~~~
\end{eqnarray} and demanding

\begin{eqnarray}
	\beta > \alpha ~, ~~ \text{and} ~~ \frac{\beta\mu_\nu^2T_\nu^2}{4\alpha V} > 1 ~, \nonumber 
\end{eqnarray} the potential 

\begin{eqnarray}\label{potential0}
\mathfrak{U} = V\left(1-\mathrm{e}^{-\alpha \phi^2/M^2_P}\right) + \frac{\mu_\nu^2 T_\nu^2}{8} \, \mathrm{e}^{-2\beta\phi^2/M_P^2} ~, 
\end{eqnarray} will have a maximum at $\phi = 0$ and two minima at

\begin{eqnarray}\label{relatmin}
	\frac{\phi_{\pm}}{M_P} = \pm \sqrt{\frac{1}{2\beta - \alpha} \ln \frac{\beta\mu_\nu^2T_\nu^2}{4\alpha V}} ~. 
\end{eqnarray} Under certain assumptions that we will shortly discuss, these minima can be considered as approximate analytic expressions for the solutions of Eq.\eqref{eqofmot}. That is, after it gets activated due to coupling with neutrinos, the field quickly adjusts itself to one of the degenerate minima (which we take for definiteness to be $\phi_+$) and then tracks this minimum. From Eq.\eqref{relatmin} one sees that as the temperature drops to

\begin{eqnarray}\label{switchoff}
	T^\star_\nu = \sqrt{\frac{4\alpha V }{\beta\mu_\nu^2}} ~, 
\end{eqnarray} the minima given by Eq.\eqref{relatmin} coalesce at $\phi=0$ and the restoration of $\mathcal{Z}_2$ symmetry takes place. Let us assume that this crossover after which the EDE disappears takes place prior to the recombination epoch, $1200 \lesssim  z \lesssim 1500 $, that is (see Eq.\eqref{temperature}),

\begin{eqnarray}\label{crossover}
	T^\star_\nu \, \simeq \, 0.2 \, \text{eV} \,-\, 0.3 \, \text{eV} ~. 
\end{eqnarray} Next we assume that the logarithm in Eq.\eqref{relatmin} is not very large and that $\beta \gg \alpha$. Under these assumptions, the small-field regime $\phi_{\pm}^2/M_P^2 << 1$ is ensured, that is,

\begin{eqnarray}
U(\phi_+) \approx \frac{\alpha V \phi_+^2}{M_P^2} ~, 
\end{eqnarray} and the slow roll condition

\begin{eqnarray}
\frac{\dot{\phi}_+^2}{2} = \frac{M_P^2H^2}{2(2\beta - \alpha)\ln \frac{\beta\mu_\nu^2T_\nu^2}{4\alpha V}} \ll \frac{\alpha V \phi_+^2}{M_P^2} = \frac{\alpha V}{2\beta - \alpha} \ln \frac{\beta\mu_\nu^2T_\nu^2}{4\alpha V} ~, \nonumber 
\end{eqnarray} is satisfied as long as\footnote{Here $\mathsf{g}_*(T_\gamma)$ denotes an effective number of relativistic degrees of freedom (at a given temperature) and $T_\nu = (4/11)^{1/3}T_\gamma$.}

\begin{eqnarray}
\frac{M_P^2H^2}{2\alpha V} = \frac{4\pi^3\mathsf{g}_*(T_\gamma)T_\gamma^4}{90 \alpha V} = \frac{5.5 \mathsf{g}_*(T_\gamma)T_\gamma^4}{\beta (\mu_\nu T^\star_\nu)^2} \ll \nonumber \\  \ln^2 \frac{\beta\mu_\nu^2T_\nu^2}{4\alpha V} = 4\ln^2 \frac{T_\nu}{T^\star_\nu} ~. \nonumber 
\end{eqnarray} To ensure that this condition holds for temperatures as high as $1$ MeV, one has to require $\beta \gg 10^{26}$. Let us note here that in view of Eq.\eqref{switchoff}, even for the lowest possible energy scale of inflation, $200$ MeV, one has to take $\beta \simeq 2.6\times 10^{31}$ in order to obtain $T_\nu^* \simeq 0.2$ eV.

In judging the accuracy of $\phi_+$ solution, let us note that by using Eq.\eqref{continuity}, the Eq.\eqref{eqofmot} can be put in the form

\begin{eqnarray}\label{phiplus}
	\frac{\mathrm{d}}{\mathrm{d}t} \left(\frac{\dot{\phi}^2}{2} + U + \rho_\nu\right) = -3H\left(\dot{\phi}^2+ \rho_\nu + p_\nu\right) ~. 
\end{eqnarray} As we have $\dot{\phi}_+^2 \ll U(\phi_+)$ and $\dot{\phi}_+^2 \ll \rho_\nu(\phi_+, T_\nu)$, we can replace Eq.\eqref{phiplus} by 

\begin{eqnarray}
&&\frac{\mathrm{d}}{\mathrm{d}t} \left( U + \rho_\nu\right) = -3H\left(\rho_\nu + p_\nu\right) ~ \Rightarrow ~ \nonumber \\&& \frac{\partial \rho_\nu}{\partial T_\nu} \dot{T}_\nu = -3H\left(\rho_\nu + p_\nu\right) ~, \nonumber 
 \end{eqnarray} which is automatically satisfied for $\rho_\nu$ and $p_\nu$ given by Eq.\eqref{simkvrive}. As for the accuracy of this solution, it can be gauged by comparing the magnitudes of $\phi_+$ and the successive term for which the equation of motion can be obtained by expanding  the right-hand side of Eq.\eqref{scalar} in power series around the $\phi_+$

 	\begin{eqnarray}
 	&&\ddot{\phi}+3H\dot{\phi} = -\mathfrak{U}''(\phi_+)  \epsilon +O(\epsilon^2) ~\Rightarrow~ \nonumber \\&& 	\ddot{\epsilon}+3H\dot{\epsilon} +\mathfrak{U}''(\phi_+)\epsilon = - \ddot{\phi}_+-3H\dot{\phi}_+ ~, \label{linearized}
 	\end{eqnarray} where $\epsilon\equiv \phi-\phi_+$. From Eq.\eqref{linearized} one sees that the condition $|\epsilon| \ll |\phi_+|$ requires the effective mass $m_{\text{eff}}^2\equiv \mathfrak{U}''(\phi_+)$ to be large. The precise criterion for the validity of applied approximation can be rewritten as

   \begin{eqnarray}\label{validity}
   \left|\frac{ \ddot{\phi}_+  +  3H\dot{\phi}_+}{\mathfrak{U}''(\phi_+)}\right| \ll |\phi_+| ~. 
   \end{eqnarray} One can immediately evaluate the left-hand side of Eq.\eqref{validity} by means of Eqs.(\ref{potential0}, \ref{relatmin})      
   
   \begin{eqnarray}\label{effectivemass}
   && m_{\text{eff}}^2(T_\nu)\equiv \mathfrak{U}''(\phi_+) = \frac{2\alpha  V}{M_P^2} \, - \, \frac{\beta  \mu_\nu^2 T_\nu^2\, \mathrm{e}^{-2\beta\phi_+^2/M_P^2}}{2M_P^2} +  \nonumber \\&& \frac{2\beta^2\phi_+^2 \mu_\nu^2 T_\nu^2\, \mathrm{e}^{-2\beta\phi_+^2/M_P^2}}{M_P^4} =  \frac{8\alpha V \ln(T_\nu/T_\nu^\star)}{M_P^2} ~,  \\ &&\dot{\phi}_+ =  -\frac{M_P H}{\sqrt{2(2\beta -\alpha)\ln T_\nu/T_\nu^*}}  ~,  \nonumber  \\ &&\ddot{\phi}_+ =  -\frac{M_P H^2}{\sqrt{2(2\beta -\alpha)\ln T_\nu/T_\nu^*}} \left(\frac{\dot{H}}{H^2} + \frac{1}{2\ln T_\nu/T_\nu^*}\right) ~. \nonumber 
   \end{eqnarray} Thus, the Eq.\eqref{validity} takes the form

   \begin{eqnarray}\label{validity condition}
   	\left| 3+ \frac{\dot{H}}{H^2} + \frac{1}{2\ln T_\nu/T_\nu^*} \right| \frac{M_P^2 H^2}{\sqrt{2} \, 8\alpha V} \ll \big(\ln T_\nu/T_\nu^*\big)^2 ~. ~~
   \end{eqnarray} In the radiation dominated universe, which is the case under consideration, $\dot{H}$ is of the order of $H^2$: $a \propto \sqrt{t}, H^2 = 1/4t^2, \dot{H} = - 1/2t^2$. On the other hand, the radiation energy density below the $\mathrm{e}^+ \mathrm{e}^-$ annihilation temperature, $0.2$ MeV, until the neutrinos become non-relativistic, can be written as\footnote{The radiation
   	content of the universe in this temperature range consists of $3$ neutrino species and the photon.}
   
   \begin{eqnarray}\label{radiation}
   \rho_r = \left(1 + 3\frac{7}{8}\left(\frac{4}{11}\right)^{4/3}\right)\rho_\gamma = 1.1 T^4_\gamma = 4.3 T^4_\nu ~, 
   \end{eqnarray} allowing the following order of magnitude estimate   
   
   \begin{eqnarray}
   	M_P^2H^2 \, \simeq \, 36 T_\nu^4 ~. 
   \end{eqnarray} Thus, roughly speaking, the validity condition \eqref{validity condition} is satisfied for $T_\nu \gtrsim T_\nu^*$ as long as the ratio $T_\nu^4/V \ll 1$. In what follows we shall need $V^{1/4}$ to be of the order of $1$\,TeV implying that $\phi_+$ is a good approximate solution and can safely be used for estimating EDE at $T_\nu\simeq 1$ eV 

\begin{eqnarray}\label{ede0}
	\rho_\phi = \frac{\alpha V}{2\beta} \ln \frac{\beta \mu_\nu^2 T_\nu^2}{4\alpha V} = \frac{(\mu_\nu T_\nu^*)^2}{4} \ln \frac{T_\nu}{T_\nu^*} ~. 
\end{eqnarray} Requiring that at $T_\nu =1$ eV the EDE comprises $10\%$ of the total energy budget of the universe, that is $5\%$ of \eqref{radiation} (because of matter-radiation equality at this temperature), one finds

\begin{eqnarray}\label{masa}
	\mu_\nu \simeq \sqrt{\frac{0.23}{0.4\times 0.2^2}} \,  \text{eV} = 3.7 \,  \text{eV} ~, 
\end{eqnarray} which is, of course, unacceptably large. On the other hand, using the parameters (\ref{massscale}, \ref{crossover}), from Eq.\eqref{ede0} one obtains just $10^{-3}\%$ of the total energy budget at $T_\nu =1$ eV. Thus, one infers that the EDE produced at the tree-level is negligibly small.

\section{Quantum fluctuations of $\phi$}

We naturally expect an additional contribution to the EDE because of quantum fluctuations of $\phi$. Recalling that $\phi$ is trapped in a minimum of $\mathfrak{U}$, which varies in time very slowly, the quantum corrections can be introduced immediately via the zero-point energy

\begin{eqnarray}\label{zero-point}
	\frac{1}{2}\int\frac{\mathrm{d}^3 q}{(2\pi)^3} \sqrt{\mathbf{q}^2 + m_{\text{eff}}^2(T_\nu)} ~, 
\end{eqnarray} where $m_{\text{eff}}^2(T_\nu)$ is given by Eq.\eqref{effectivemass}. Choosing the normalization such that the quantum correction \eqref{zero-point} vanishes when $m_{\text{eff}}^2 = 0$, that is at $T_\nu = T_\nu^*$, one can express the result in terms of the one-loop effective potential (see the next section)

	\begin{eqnarray}\label{quantum}
	&&\frac{1}{2}\int\frac{\mathrm{d}^3q}{(2\pi)^3} \left(\sqrt{\mathbf{q}^2+m_{\text{eff}}^2(T_\nu)} -q\right)    =  \frac{1}{2}	\int\frac{\mathrm{d}^4 q}{(2\pi)^4} \times \nonumber \\&&  \ln \frac{q^2+m_{\text{eff}}^2(T_\nu)}{q^2}  =   \frac{m_{\text{eff}}^4(T_\nu)}{64\pi^2}\left(\ln\frac{m^2_{\text{eff}}(T_\nu)}{\mu^2} + \frac{1}{2} \right)~,
	\end{eqnarray} where $\mu$ stands for the renormalization scale. Let us first set $\mu$ by the neutrino mass scale $\mu_\nu$. Then for obtaining the needed amount of EDE at $T_\nu =1$ eV, one has to take

	\begin{eqnarray}\label{infenscale}
		V \, \simeq \,  10^{-2} \,\text{eV}^2 \times M_P^2 \approx 10^{54}\, \text{eV}^4 ~.  
	\end{eqnarray} In view of Eq.\eqref{switchoff}, this value of inflation energy scale, in conjunction with the requirement $T_\nu^* \simeq 0.2$ eV, leads to $\beta \sim 10^{59}$. It is curious to note that for this value of $\beta$ - the mass scale $M_P^2/\beta$ appearing in the coupling function \eqref{example1} is of the order of $\mu_\nu^2$. From Eq.\eqref{infenscale}) one sees that the mass of the scalar field $m_\phi=\sqrt{2\alpha V/M_P^2}$ is quite close to $\mu_\nu$. Therefore, setting $\mu$ by the mass of the scalar field, one would have almost the same value of $V$

	\begin{eqnarray}\label{infenscale2}
		V \simeq 10^{-1} \, \text{eV}^2\times M_P^2 ~.  
	\end{eqnarray}
	
It is worth paying attention that because of interaction with CNB, there will be quantum and also thermal corrections to $\mathfrak{U}(\phi)$. These corrections are usually evaluated in terms of the Coleman-Weinberg effective potential \cite{Coleman:1973jx} by integrating the fermionic degrees of freedom out of the partition function \cite{Kapusta:2006pm, Bailin:1986wt}. We shall now address this question.

\section{Radiative and thermal corrections coming from neutrino sector}

Using a functional integral representation of the partition function and doing the Matsubara sums \cite{Kapusta:2006pm, Bailin:1986wt}, one can write the thermal and one-loop quantum corrections to the scalar field potential $\mathfrak{U}(\phi)$ as  

 \begin{eqnarray}\label{fermion}
- 3 \int\frac{\mathrm{d}^3q}{(2\pi)^3} \left(\sqrt{\mathbf{q}^2+m_\nu^2(\phi)} -q\right) -  ~~~~~~~~~~~~ \nonumber \\   6T_\nu\int\frac{\mathrm{d}^3q}{(2\pi)^3} \ln \frac{1+\mathrm{e}^{-\left.\sqrt{\mathbf{q}^2+m_\nu^2(\phi)}\right/T_\nu}}{1+\mathrm{e}^{- q/T_\nu}}   ~. 
\end{eqnarray} Here we have taken into account the presence of three species of neutrinos and have normalized the expression \eqref{fermion} in such a way that it vanishes when $m_\nu^2(\phi) = 0$. The second term in Eq.\eqref{fermion} corresponding to the thermal corrections is convergent, while the first integral, which represents one-loop quantum correction, is divergent. For evaluating quantum correction, one can put the integral in the standard manner in a Wick rotated $4$-dimensional form and then exploit a $4$-momentum cutoff 

\begin{eqnarray}&& \int\frac{\mathrm{d}^3q}{(2\pi)^3} \left(\sqrt{\mathbf{q}^2+m_\nu^2} -q\right) =  \int \mathrm{d}^4 q \,  \ln \frac{q^2+m_\nu^2(\phi)}{q^2} = \nonumber \\&&  2\pi^2\int_0^{~\, q_c} {q^3dq\ln\left(\frac{q^2+m_\nu^2}{q^2}\right)} =  \frac{\pi^2q_c^4}{2}\ln\left( \frac{q_c^2+m_\nu^2(\phi)}{q_c^2}\right) \,+ \nonumber \\&&
\frac{\pi^2m_\nu^2(\phi)q_c^2}{2} \,-\, \frac{\pi^2m_\nu^4}{2}\ln\left(\frac{q_c^2+m_\nu^2(\phi)}{m_\nu^2(\phi)} \right) = \pi^2 m_\nu^2q_c^2  \, -  \nonumber \\&& \frac{\pi^2 m_\nu^4}{2} \left(\ln\frac{\mu^2}{m_\nu^2} + \ln\frac{q_c^2}{\mu^2}+\frac{1}{2}\right) + O\left(\frac{m_\nu^6}{q_c^2}\right)  ~. \nonumber 
\end{eqnarray} Omitting here the terms vanishing when $q_c\to\infty$ and substituting an expression of $m_\nu$ from Eq.\eqref{example1}, one obtains a power series expansion of the divergent terms in the even powers of $\phi$. All these even powers of $\phi$ can be found in the tree-level effective potential \eqref{example1} and correspondingly all divergences can be reabsorbed in the corresponding coupling parameters. This procedure of renormalization results in the logarithmic term\footnote{The expression \eqref{quantum} is obtained in much the same way.}

\begin{eqnarray}\label{fermionco}
	\int \mathrm{d}^4 q \,  \ln \frac{q^2+m_\nu^2(\phi)}{q^2} = -\frac{\pi^2 m_\nu^4}{2} \left(\ln\frac{\mu^2}{m_\nu^2} +\frac{1}{2}\right) ~, 
\end{eqnarray} which depends on the renormalization scale $\mu$. One can naturally set this scale by the mass of the neutrino $\mu_\nu$. In addition, as a part of the renormalization procedure, we add the term $\pi^2 m_\nu^4(0)/4$ to Eq.\eqref{fermionco} in order to avoid the cosmological constant (of the order of $\mu_\nu^4$) left over after the symmetry restoration. This way one obtains the potential

\begin{eqnarray}\label{fermionuli}
\mathfrak{U}_Q =  V\left(1- \mathrm{e}^{-\alpha\phi^2/M_P^2}\right) +  \frac{\mu_\nu^2 T_\nu^2}{8} \mathrm{e}^{-2\beta\phi^2/M_P^2} + \nonumber \\  \frac{3 \mu_\nu^4 \,\mathrm{e}^{-4\beta\phi^2/M_P^2}}{32\pi^2}\left( \frac{2\beta\phi^2}{M_P^2} + \frac{ 1}{2} \right) - \frac{3\mu_\nu^4}{64\pi^2} ~. 
\end{eqnarray} The quantum correction in the second line of Eq.\eqref{fermionuli} (which is negative for $|\phi| > 0$ as it should be in view of Eq.\eqref{fermion}) does not affect the solution \eqref{relatmin} for $T\gtrsim T_\nu^* $. To see it let us first evaluate

\begin{eqnarray}\label{compound}
	\frac{\alpha V \phi_+^2}{M_P^2} +  \frac{\mu_\nu^2 T_\nu^2}{8} \mathrm{e}^{-2\beta\phi_+^2/M_P^2} =    \frac{(\mu_\nu T_\nu^*)^2}{8} \left(2 \ln\frac{T_\nu}{T_\nu^*}	+  1\right) ~. ~~~~~~ 
\end{eqnarray} As to the quantum correction, one finds

      \begin{eqnarray}\label{qshestsoreba}
      	\frac{3 \mu_\nu^4 \,\mathrm{e}^{-4\beta\phi_+^2/M_P^2}}{32\pi^2}\left( \frac{2\beta\phi_+^2}{M_P^2} + \frac{ 1}{2} \right)  =   \frac{3 \mu_\nu^4}{32\pi^2} \left(\frac{T_\nu^*}{T_\nu}\right)^4 \times \nonumber \\  \left(2\ln\frac{T_\nu}{T_\nu^*} +\frac{1}{2}\right) = \frac{\mu_\nu^4}{105}  \left(\frac{T_\nu^*}{T_\nu}\right)^4 \left(2\ln\frac{T_\nu}{T_\nu^*} +\frac{1}{2}\right)   ~. ~~  
      \end{eqnarray} Taking for instance $T_\nu = 2T_\nu^*$ and noting that

      \begin{eqnarray}
      	\left.\frac{\mu_\nu^4}{105}\right/ \frac{(\mu_\nu T_\nu^*)2}{8}  \approx 0.005 ~, \nonumber 
      \end{eqnarray} the Eq.\eqref{qshestsoreba} becomes by a factor $0.0003$ smaller than Eq.\eqref{compound}. By increasing the temperature further, this ratio decreases as $(T_\nu^*/T_\nu)^4$. For the temperature $1 \, \text{eV} (= 5T_\nu^*)$ the contribution of the quantum correction \eqref{qshestsoreba} to the tree-level EDE density \eqref{compound} is suppressed by a factor $0.000008$ and can be safely ignored. Interestingly enough, the correction in Eq.\eqref{fermionuli} does not affect the crossover temperature. Namely, the crossover temperature $T_\nu^*$ is determined by setting $\phi = 0$ in the extremum condition

      \begin{eqnarray}
      	2\alpha V \mathrm{e}^{-\alpha \phi^2/M_P^2} = \frac{\beta \mu_\nu^2 T_\nu^2 \mathrm{e}^{-2\beta \phi/M_P^2}}{2}  + \frac{3\beta^2 \mu_\nu^2\phi^2}{4\pi^2M_P^2} \mathrm{e}^{-4\beta\phi^2/M_P^2} ~, \nonumber 
      \end{eqnarray} and is thus tantamount to the tree-level condition.

Hawing observed that quantum corrections due to coupling with CNB do not affect the theory, let us turn to the thermal corrections. In particular we are interested in evaluating thermal corrections in the high-temperature regime: $T_\nu \gtrsim \mu_\nu$. It can be done straightforwardly from Eq.\eqref{fermion} \cite{Kapusta:2006pm, Bailin:1986wt} resulting in the approximate expression $m_\nu^2 T_\nu^2/8$ \cite{Kapusta:2006pm, Bailin:1986wt}. This contribution makes the last term in Eq.\eqref{largeT} twice bigger and thus leads to the replacement $\mu_\nu^2\to 2\mu_\nu^2$ in the subsequent equations. This change does not affect anything significantly.

\section{Discussion and conclusions}

In the present model EDE is sourced by the quantum fluctuations of $\phi$ and the magnitude of this contribution is controlled by the energy scale of inflation. The role of neutrinos is that they activate inflaton at the early stage of Big-Bang leading to the dynamical braking of $\mathcal{Z}_2$ symmetry and then, after the scalar field gets trapped in a local minimum of the effective potential $\mathfrak{U}(\phi)$, they ensure slow roll regime for the field. The reason why the model cannot explain EDE at the tree-level lies in the fact that the model is basically controlled by two parameters $\mu_\nu$ and $T_\nu^*$ both of which are smaller than $1$\,eV (see Eqs.(\ref{massscale}, \ref{crossover})). In particular, in the high-temperature regime, $\rho_\nu$ gets correction due to neutrino masses, which is a constant fraction to the energy budget

\begin{eqnarray}
	\frac{m_\nu^2(\phi_+)T_\nu^2}{8} = \frac{(\mu_\nu T_\nu^*)^2}{8}  ~, \nonumber 
\end{eqnarray} while the inflaton energy density (which grows exceedingly slowly with $T_\nu$) is given by

\begin{eqnarray}
\rho_\phi = \frac{(\mu_\nu T_\nu^*)^2}{8} \ln\left(\frac{T_\nu}{T_\nu^*}\right)^2 ~.    \nonumber 
\end{eqnarray} Thus, one can readily estimate the order of magnitude of EDE as $(\mu_\nu T_\nu^*)^2$. This explains why we need large neutrino masses \eqref{masa}, exceeding the cosmological bound $\sum m_\nu \lesssim 0.15$\,eV (see section "neutrino properties" in \cite{ParticleDataGroup:2020ssz}), for producing a needed amount of EDE. It is worth paying attention that the energy density excess of inflaton-neutrino compound as compared to the case when neutrino masses are independent of $\phi$   

\begin{eqnarray}
\frac{(\mu_\nu T_\nu^*)^2}{8}  \left( \ln\left(\frac{T_\nu}{T_\nu^*}\right)^2 + 1  \right) - \frac{(\mu_\nu T_\nu)^2}{8}  ~, \nonumber 
\end{eqnarray} is negative for $T_\nu > T_\nu^*$.

Despite the fact that the tree-level effective potential fails to provide a sufficient amount of EDE - it may provide effective mass \eqref{effectivemass} needed for quantum corrections to fit the required EDE density. Remarkably enough, this is achieved for relatively low energy scale of inflation, Eqs.(\ref{infenscale}, \ref{infenscale2}), of the order of electroweak energy scale. A subtle point in properly evaluating quantum corrections is that one can perform formally a renormalization procedure which allows to avoid cutoff dependence of the result. It is understood that first one integrates fermionic degrees of freedom out of the partition function. This way, one observes that the emerging quantum and thermal corrections to the effective potential (which are of the order of $\hbar$) do not affect the minimum $\phi_+$ significantly. In view of the fact that $\phi_+$ varies very slowly, the correction to the vacuum $\mathfrak{U}_Q(\phi_+)$ due to quantum fluctuations of $\phi$ is just a zero-point energy. As we are interested in corrections that are of order $\hbar$, in evaluating effective mass \eqref{effectivemass} one should use the tree-level expression $\mathfrak{U}(\phi_+)$. Forgetting about the renormalization, one could try to evaluate the zero-point energy by exploiting the $3$-momentum cutoff 

\begin{eqnarray}\label{threemomentum}
\int_0^{q_c}\frac{\mathrm{d}q }{4\pi^2} \,  q^2\left(\sqrt{q^2+m_{\text{eff}}^2(T_\nu)} -q\right)    = \frac{m_{\text{eff}}^2(T_\nu) q_c^2}{16\pi^2}  + \nonumber \\ \frac{m_{\text{eff}}^4(T_\nu)}{64\pi^2}\left(\ln\frac{m^2_{\text{eff}}(T_\nu)}{4q_c^2} + \frac{1}{2} \right) +O\left(\frac{m_{\text{eff}}^6(T_\nu)}{q_c^2}\right) ~, ~~
\end{eqnarray} and then omitting the terms violating Lorentz invariance\footnote{What is meant here is that in the expression of zero-point energy, neither quartic nor quadratic term coming from the 3-momentum cutoff respects the Lorentz invariance \cite{Akhmedov:2002ts, Ossola:2003ku, Koksma:2011cq}.} \cite{Akhmedov:2002ts, Ossola:2003ku, Koksma:2011cq}. But in order to suppress the terms 

\begin{eqnarray}
m_{\text{eff}}^4(T_\nu) \left(\frac{m_{\text{eff}}(T_\nu)}{q_c}\right)^j ~, ~~j = 2, 4, 6, \ldots ~, \nonumber 
\end{eqnarray} in Eq.\eqref{threemomentum} the cutoff should be large enough, which in turn implies that the logarithmic term we are interested in becomes negative and therefore it alone cannot stand for the integral \eqref{threemomentum} as it is clearly positive-definite.

For our discussion the gravitational corrections related 1) to the graviton loops and 2) to the presence of curved background can be safely ignored. Namely, the radiative corrections to the potential that arises due to one-loop graphs involving gravitons can be written in terms of \cite{Smolin:1979ca}

\begin{eqnarray}\label{graviton}
	\frac{\mathfrak{U}(\phi)}{M_P^4} ~, ~~ \frac{\mathfrak{U}'(\phi)}{M_P^3}  ~,~~ \frac{\mathfrak{U}''(\phi)}{M_P^2} ~. 
\end{eqnarray} In view of the fact that (here we use $\beta \simeq 10^{58}$)

\begin{eqnarray}
	\phi_+ \simeq \mu_\nu \sqrt{\ln\frac{T_\nu}{T_\nu^*}} \ll M_P ~, \nonumber 
\end{eqnarray} one infers without much ado that all terms in Eq.\eqref{graviton} evaluated at $\phi_+$ are strongly suppressed. One may also wonder about the corrections because of curved background space-time. In curved space-time, the effective potential can be regarded as being expanded in powers of the curvature \cite{Parker:2009uva}. Restricting ourselves to the linear order in curvature,

\begin{eqnarray}
R = 6\left[\frac{\dot{a}^2}{a^2} + \frac{\ddot{a}}{a}\right] ~, \nonumber 
\end{eqnarray} one finds that for the radiation dominated universe, $a\propto t^{1/2}$, the curvature is zero and thus the corrections become strongly suppressed.

One more important issue worth paying attention is that in most cases the mass varying neutrino models predict the clumpy structure of CNB \cite{Afshordi:2005ym}. However, these clumps arise as cooling instabilities where the cooling temperature is set by the neutrino mass scale and as the mass variation of neutrinos in the present model stops before CNB enters the non-relativistic regime, the model is free of this instability.

A particular model we have considered can be generalized straightforwardly to a broad class of "good" inflationary potentials. Under this term we understand potentials having a plateau, which provides a slow-roll regime, and a minimum at $\phi=0$ around of which the field starts to oscillate after it exits the slow-roll regime \cite{Linde:2018hmx, Kallosh:2019hzo}. The potentials with plateau provide perfect conditions for the slow-roll inflation as the field rolling down will arrive at the attractor trajectory from a very wide range of initial conditions. Interestingly enough, in the above discussion one can straightforwardly use the broad class of inflationary potentials derived in \cite{Kallosh:2013hoa} as a result of spontaneously broken conformal symmetry. For instance, one can readily generalize our tree-level discussion to the T-model potentials \cite{Carrasco:2015rva}

\begin{eqnarray}
U\left(\tanh^2 \frac{\phi}{\sqrt{6\alpha M_P^2}}\right)  ~, \nonumber 
\end{eqnarray} (with the same $\phi$-$\nu$ coupling) as they are closely analogous to what we have considered. However, in the cases when cutoff dependence of quantum corrections can not be evaded (even by the formal procedure we have exploited), the model becomes complicated by the ambiguity in estimating the proper size of one-loop corrections. In particular it is caused by the presence of $m^2q_c^2$ term in quantum corrections. In general, this term is hard to control in non-renormalizable theories that cannot be put in the framework of an effective field theory. In much the same way, in our discussion one could consider E-model potentials \cite{Carrasco:2015rva} for the inflaton field. 

In view of the above discussion, the obvious question to ask is - what if one considers a strictly renormalizable derived model by dropping all higher order terms in the field? Such model can be derived from \eqref{example1} by truncating $m_\nu$ at the quadratic order. As we are interested in a high-temperature regime, $T_\nu \gtrsim \mu_\nu$, the effective potential takes the form

\begin{eqnarray}
\mathfrak{U} = \frac{\alpha V \phi^2}{M_P^2} + \frac{\mu_\nu^2T_\nu^2}{8} \left(1 - \frac{\beta \phi^2}{M_P^2}\right)^2 ~. \nonumber 
\end{eqnarray} This potential has the minima at

\begin{eqnarray}
\frac{\beta \phi^2_{\pm}}{M_P^2} =  1 \,-\, \frac{4\alpha V}{\beta \mu_\nu^2 T_\nu^2}  =   1 \,-\, \left(\frac{T_\nu^*}{ T_\nu}\right)^2 ~,
\end{eqnarray} where we have used a crossover temperature at which the symmetry restoration takes place  

\begin{eqnarray}\label{kritische Temperatur}
T_\nu^* = \sqrt{\frac{4\alpha V}{\beta \mu_\nu^2}} ~. 
\end{eqnarray} Even if we take an inflationary energy scale as low as possible, $V\simeq 200$ MeV, one should take $\beta$ quite large, of the order of $10^{38}$, in order to obtain $T_\nu^* \simeq 0.2$ eV. The slow roll condition

\begin{eqnarray}
&&\frac{\dot{\phi}_+^2}{2} =  \frac{M_P^2H^2}{2\beta \big[1 - (T_\nu^*/T_\nu)^2\big]} \left(\frac{T_\nu^*}{T_\nu}\right)^4 = \nonumber \\&&  \frac{4\pi^3\mathsf{g}_*(T_\gamma)T_\gamma^4}{90 \beta \big[1 - (T_\nu^*/T_\nu)^2\big]} \left(\frac{T_\nu^*}{T_\nu}\right)^4 = \frac{15.6\pi^3\mathsf{g}_*(T_\gamma)(T_\nu^*)^4}{90 \beta \big[1 - (T_\nu^*/T_\nu)^2\big]} \ll \nonumber \\&&  \frac{\alpha V}{\beta} \left[  1 \,-\, \left(\frac{T_\nu^*}{ T_\nu}\right)^2\right] = \frac{(\mu_\nu T_\nu^*)^2}{4} \left[  1 \,-\, \left(\frac{T_\nu^*}{ T_\nu}\right)^2\right] ~, \nonumber 
\end{eqnarray} is tantamount to $\beta \gg 10^3$ and is thus satisfied with a great accuracy. We can consider $\phi_+$ as a good approximate solution and evaluate EDE as

\begin{eqnarray}
\rho_\phi = \frac{\alpha V}{\beta} \left[  1 \,-\, \left(\frac{T_\nu^*}{ T_\nu}\right)^2\right] = \frac{(\mu_\nu T_\nu^*)^2}{4} \left[  1 \,-\, \left(\frac{T_\nu^*}{ T_\nu}\right)^2\right] ~. ~~~~~~
\end{eqnarray} Requiring that at $T_\nu =1$ eV the EDE comprises $10\%$ of the total energy budget of the universe, one gets again unacceptably large neutrino mass 

\begin{eqnarray}
\mu_\nu \simeq \left(\frac{0.46 \, \text{eV}^4}{0.04 \, \text{eV}^2}\right)^{1/2} \approx 3.4 \, \text{eV} ~. \nonumber 
\end{eqnarray} Thus, as in the previous case, the tree-level EDE is merely negligible. The correction to EDE density due to quantum fluctuations of $\phi$ can be evaluated as

\begin{eqnarray}\label{gantoleba}
\big(\mathfrak{U}''(\phi_+)\big)^2 \ln\frac{\mathfrak{U}''(\phi_+)}{\mu} = \frac{\beta^2\mu_\nu^4\Big(T_\nu^2-(T_\nu^*)^2\Big)^2}{M_P^4}\times \nonumber \\ \ln2\left[\left(\frac{T_\nu}{T_\nu^*}\right)^2 - 1\right] = \frac{16\alpha^2 V^2\Big(T_\nu^2-(T_\nu^*)^2\Big)^2}{(M_PT_\nu^*)^4} \times \nonumber \\ \ln2\left[\left(\frac{T_\nu}{T_\nu^*}\right)^2 - 1\right]  ~, ~~~~~
\end{eqnarray} where we have set the renormalization scale $\mu$ by the mass of scalar field, $m_\phi = 2\alpha V/M_P^2$ and used the relation \eqref{kritische Temperatur}. Thus, one can again obtain a needed amount of EDE at the expense of $V$ but it increases with temperature faster than $T_\nu^4$ that makes this model obviously inconsistent with the early time cosmology.

\begin{acknowledgments} The useful discussions with Zurab Kepuladze, Bharat Ratra and Vakhtang (Vato) Tsintsabadze are kindly acknowledged. Author is also indebted to Gennady Chitov and Tina Kahniashvili for useful comments. The work was supported in part through funds provided by the Rustaveli National Science Foundation of Georgia under Grant No. FR-19-8306 and by the Ilia State University under the Institutional Development Program.
\end{acknowledgments}

\nocite{*}
\bibliographystyle{apsrev4-1}
\bibliography{Literature}

\end{document}